\documentclass[preprint2]{aastex}    

\usepackage{xcolor}

\shorttitle{Double nucleus galaxies}
\shortauthors{Mezcua et al.}

\begin{document}

\title{PHOTOMETRIC DECOMPOSITION OF MERGERS IN DISK GALAXIES}

\author{M.~Mezcua\altaffilmark{1,2,3}, A.P.~Lobanov\altaffilmark{3,4}, E.~Mediavilla\altaffilmark{1,2}, and M. Karouzos\altaffilmark{5}}

\affil{$^{1}$Instituto de Astrof\'isica de Canarias (IAC), E-38200 La Laguna, Tenerife, Spain;  mmezcua@iac.es}
\affil{$^{2}$Universidad de La Laguna, Dept. Astrof\'isica, E-38206 La Laguna, Tenerife, Spain} 
\affil{$^{3}$Max Planck Institute for Radio Astronomy, Auf dem H\"ugel 69, D-53121 Bonn, Germany}
\affil{$^{4}$Institut f\"ur Experimentalphysik, Universit\"at Hamburg, Luruper Chausse 149, D-22761 Hamburg, Germany}
\affil{$^{5}$Center for the Exploration of the Origin of the Universe, Seoul National University, Seoul 151-742, Korea}



\begin{abstract}
 Several observational studies and numerical simulations suggest that mergers must contribute to the evolution of galaxies; however, the role that they play is not yet fully understood. In this paper we study a sample of 52 double nucleus disk galaxies that are considered as candidates for a minor merger event. The luminosity of each of the nuclei and their relative separation are derived from a multi-component photometric fit of the galaxies in the Sloan Digital Sky Survey optical images. We find that the nuclei in most of the sources have projected separations $\leq$ 4 kpc. The ratio of nuclear luminosities indicates that most of the systems are likely in the \textit{coalescence} stage of a major merger. This is supported by the existence of a single galaxy disk in 65\% of the systems studied and the finding of a correlation between nuclear luminosity and host luminosity for the single-disk systems: those sources fitted with as single disk are in a more evolved stage of the merger and present an enhancement of the nuclear luminosity compared to the double-disk systems, as expected from simulations of galaxy mergers.
Finally, we identify a sample of 19 double nucleus disk galaxies in which the two nuclei are physically separated by $\leq1$ kpc and constitute thus a sample of sub-kpc binary active galactic nucleus candidates.
 \end{abstract}


\section{Introduction}
A galaxy merger,
understood as a pair of galaxies that are gravitationally bound and
whose orbits will dynamically decay until their nuclei merge (\citealt{1972ApJ...178..623T}), is a
process that lasts a few Gyr (e.g., 1--3 Gyr for major mergers with
mass ratios $>$1:3) and passes through different stages
(e.g., \citealt{2008MNRAS.391.1137L}). 
Numerical simulations show that in major mergers of disk galaxies there
is an enhancement of star formation and active galactic nucleus (AGN) activity during the coalescence
stage, when the nuclei merge. This occurs 1.5--2 Gyr after the pre-merger stage (e.g., \citealt{2005MNRAS.361..776S}; \citealt{2008MNRAS.391.1137L}) depending on the disks mass and orbital configuration of the system (e.g., \citealt{2008MNRAS.391.1137L}). A time delay of about 0.1--0.5 Gyr may occur between the observed peak of nuclear starburst and the AGN-triggered activity, either due to intrinsic effects related to the AGN triggering mechanism or due to obscuration of the AGN emission by the surrounding gas and dust (e.g., \citealt{2005MNRAS.361..776S}; \citealt{2008ApJS..175..356H}; \citealt{2010ApJ...714L.108S}). Possible observational evidence for this phenomenon has been suggested in some radio galaxies (e.g., \citealt{2006A&A...454..125E}; \citealt{2012A&A...544A..36M}).
The most pronounced morphological distortions take place also during the coalescence stage (e.g., \citealt{2005Natur.433..604D}; \citealt{2008MNRAS.391.1137L}) and are expected to vanish within about 0.5 Gyr after the nuclei merge, in the post-merger stage (e.g., \citealt{2008MNRAS.391.1137L}), when a single remnant galaxy is formed.

In the hierarchical galaxy formation models (e.g.,
\citealt{1993MNRAS.262..627L,1994MNRAS.271..676L,2003ApJ...582..559V}),
galaxies grow in a $\Lambda$ cold dark matter ($\Lambda$CDM) universe
in a `bottom-up' way through multiple mergers. However, the role of mergers in the growth of the stellar mass and size of local massive galaxies is not yet well understood, especially in terms of its empirical aspect, while alternative processes such as secular evolution (e.g., cold gas accretion) can naturally explain the formation of disk galaxies and pseudobulges (e.g., \citealt{2004ARA&A..42..603K}; \citealt{2012arXiv1212.1463S}).

Major mergers have been shown to be responsible for only $\sim20\%$ of the
mass growth of massive galaxies at $z < 1$ (e.g., \citealt{2009A&A...498..379D}; \citealt{2010ApJ...710.1170L}), which is a significant but not dominant fraction. Therefore, other mechanisms such as cold gas accretion or minor mergers must contribute to the mass growth of galaxies.
Minor mergers can lead to a significant increase in mass (e.g., from $\sim$25\% at $z \lesssim 1$ to a factor $\sim$2 at $z < 3$; e.g., \citealt{2011A&A...530A..20L}; \citealt{2012ApJ...747...34B}) and have been found to be more efficient at increasing galaxy radii than major mergers. A mass increase due to minor mergers of a factor two can lead to a size increase of up to a factor four in radii in massive galaxies at $z < 3$ (e.g., \citealt{2012ApJ...747...34B}). The factor of size growth in massive galaxies due to minor mergers is found to range between 2 and 5 at $z \lesssim 1$--2 in both observational (e.g., \citealt{2005ApJ...626..680D}; \citealt{2006ApJ...650...18T}; \citealt{2008ApJ...687L..61B}; \citealt{2012MNRAS.422L..62C}; \citealt{2013MNRAS.428.1715H}) and theoretical (e.g., \citealt{2007A&A...476.1179B}; \citealt{2009ApJ...699L.178N}; \citealt{2012ApJ...744...63O}) studies.

The contribution of minor mergers to star formation is also significant (e.g., $\geq$35\% over cosmic time; \citealt{2014MNRAS.437L..41K}) and evidence of minor merging events have been often reported in the literature in the form of recent star formation in early-type galaxies (e.g., \citealt{2007ApJS..173..619K,2009MNRAS.394.1713K}; \citealt{2011MNRAS.411L..21F}).

Additional direct observations of galaxy mergers and especially of the much less empirically explored minor mergers are needed in order to constrain the galaxy evolutionary models.
Several catalogs and surveys of merging
galaxies with double nuclei exist (e.g., \citealt{1991AJ....101.2034M};
\citealt{1993ApJS...85...27M}; \citealt{2004AJ....128...62G};
\citealt{2007ApJ...666..212D}; \citealt{2010ApJ...709.1067B}; \citealt{2011MNRAS.410..166L}; \citealt{2012A&A...539A..45L}; \citealt{2012ApJ...746L..22K} and references therein), while many
studies have been aimed at detecting binary nuclei inferred from AGN properties such as double-peaked
emission lines (e.g., \citealt{2010ApJ...708..427L}a, \citeyear{2010ApJ...715L..30L}b;
\citealt{2011ApJ...738L...2M}; \citealt{2014MNRAS.437...32W}) or peculiar jet structures (e.g.,
\citealt{2005A&A...431..831L}; \citealt{2008A&A...483..125R};
\citealt{2011A&A...527A..38M,2012A&A...544A..36M}).  

With the aim of studying the role of minor merger processes in disk galaxies in more detail, we analyze in this paper a small sample of double nucleus disk galaxies classified as minor mergers by \cite{2004AJ....128...62G}.
A photometric point spread function (PSF) fitting is used to estimate the brightness of
each of the nuclei and the distance between them in the optical images provided by the
Sloan Digital Sky Survey (SDSS)\footnote{Sloan Digital Sky Survey, www.sdss.org}. Based on these
fits, we obtain the nuclear luminosities  and the separation between the two nuclei in these systems and study their correlation with the host galaxy.

The sample of galaxies analyzed is
presented in Section~\ref{sec:sample}. The description of the PSF fitting technique and of the
analysis performed is explained in Section~\ref{analysis}. The
results obtained are presented in Section~\ref{results} and
discussed in Section~\ref{discussion}. Final conclusions are given in Section~\ref{conclusions}.

Through this paper, we assume a $\Lambda$ CDM cosmology with parameters $H_\mathrm{0} = 73$ km s$^{-1}$ Mpc$^{-1}$, $\Omega_{\Lambda}=0.73$, and $\Omega_\mathrm{m}=0.27$.

\section{The sample}
\label{sec:sample}
The sources analyzed are drawn from a catalog of double nucleus disk
galaxies suggested as candidates for minor merger events
(\citealt{2004AJ....128...62G}). The catalog comprises 107
double nucleus galaxies selected from older catalogs under the
following three requirements: (1) galaxies with redshift $z < 0.05$ (or
apparent \textit{B} magnitude $m_\mathrm{B} < 18$); (2) galaxies
showing disk-like morphology (i.e., elliptical and cD galaxies are
excluded) to avoid the inclusion of major merger remnants and include
only likely minor merger events; (3) galaxies not exhibiting strong
tidal distortions or tails, as these features are expected in major
merger events (\citealt{2004AJ....128...62G}).

\begin{figure*}
  \includegraphics[scale=0.65]{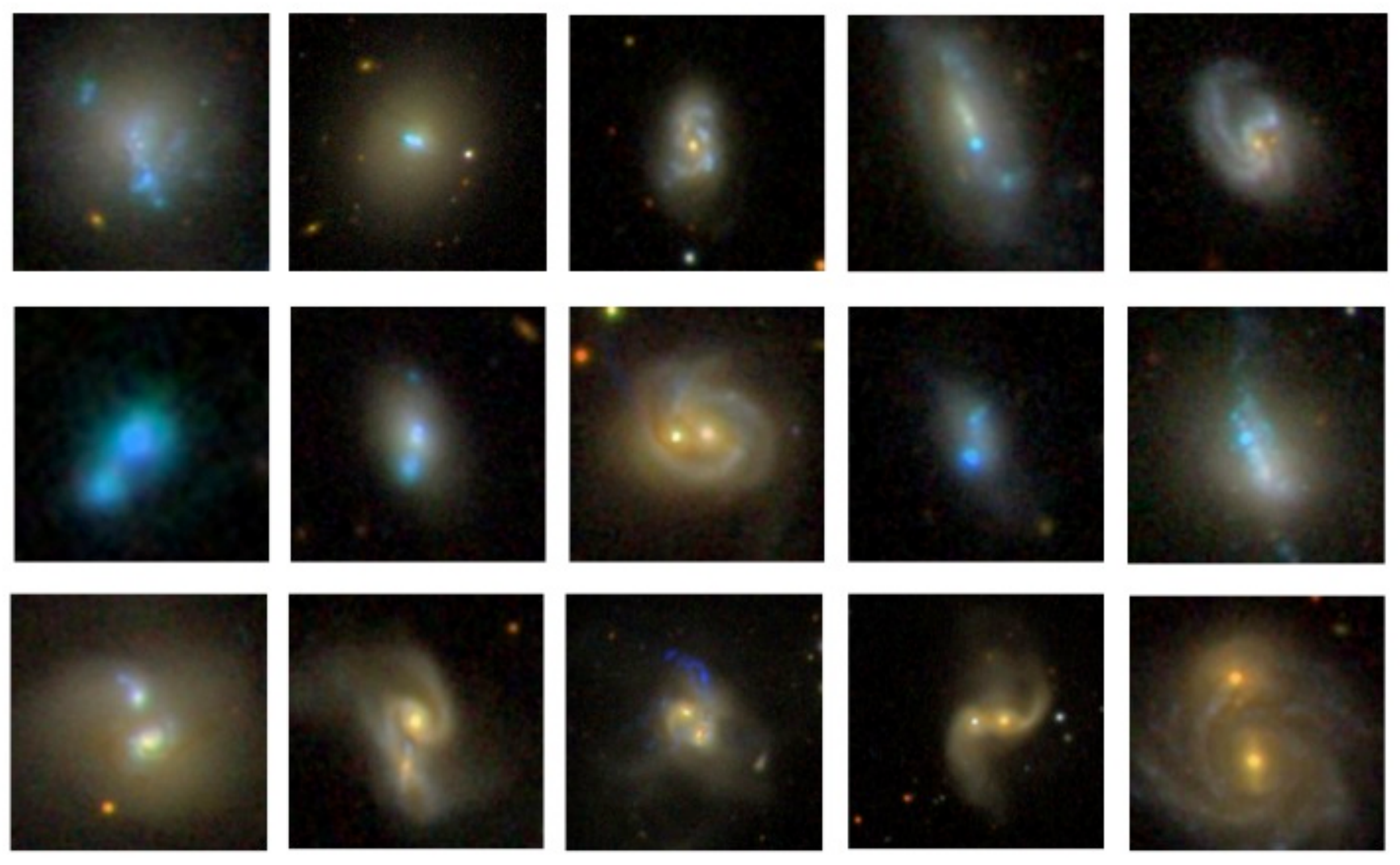}
  \protect\caption[Composite JPG]{From top to bottom, left to right: color \textit{gri} SDSS DR8 images of the sources NGC\,5058, NGC\,3773, Mrk\,1114, Mrk\,712, Mrk\,721, Mrk\,116, Mrk\,104, NGC\,3758, Mrk\,1263, NGC\,7468, NGC\,5860, Mrk\,423, NGC\,5256, Mrk\,212, and MCG +00-12-073. Sources are ordered in ascending nuclear separation. The field of view is different for each object (i.e., 25 arcsec $\times$ 25 arcsec, 51 arcsec $\times$ 51 arcsec, and 100 arcsec $\times$ 100 arcsec) so that the morphological type of the host galaxy can be appreciated. \\(A color version of this figure is available in the online journal.) \label{composite_jpg}}
\end{figure*}

Of the 107 double nucleus disk galaxies, 60 are found to have imaging
data in the SDSS Data Release 8
(SDSS DR8
hereafter; \citealt{2011AJ....142...72E}, and references therein). Images of some of these sources in {\em gri} colors are shown in Figure~\ref{composite_jpg}.
FITS images of the 60 objects in the \textit{u},
\textit{g}, and \textit{r} bands (centered at 3551\AA, 4686\AA, and
6166\AA, respectively) are retrieved by coordinates from the Science
Archive Server (SAS) of the SDSS with a size of 2048 $\times$ 1489 pixels and a pixel
scale of 0$^{\prime\prime}$.396 pixel$^{-1}$.  The retrieved images are
calibrated and have an applied sky-subtraction appropriate for large objects
(see \citealt{2011AJ....142...72E} for a more detailed description).

For the objects MCG +06-07-020 and Mrk\,1341, the coordinates reported in \cite{2004AJ....128...62G} do not provide the SDSS field image where these sources should be located. The corrected coordinates are found in NED\footnote{NASA/IPAC Extragalactic Database.} by name and are, for MCG +06-07-020: R.A.(J2000)=02$^\mathrm{h}$55$^\mathrm{m}$29$^\mathrm{s}$, decl.(J2000)=36$^{\circ}$12$^{\prime}$0$^{\prime\prime}$.2 and for Mrk 1341: \\R.A.(J2000)=13$^\mathrm{h}$00$^\mathrm{m}$59$^\mathrm{s}$, decl.(J2000)=--00$^{\circ}$01$^{\prime}$39$^{\prime\prime}$.

\begin{figure*}
  \includegraphics[scale=0.65]{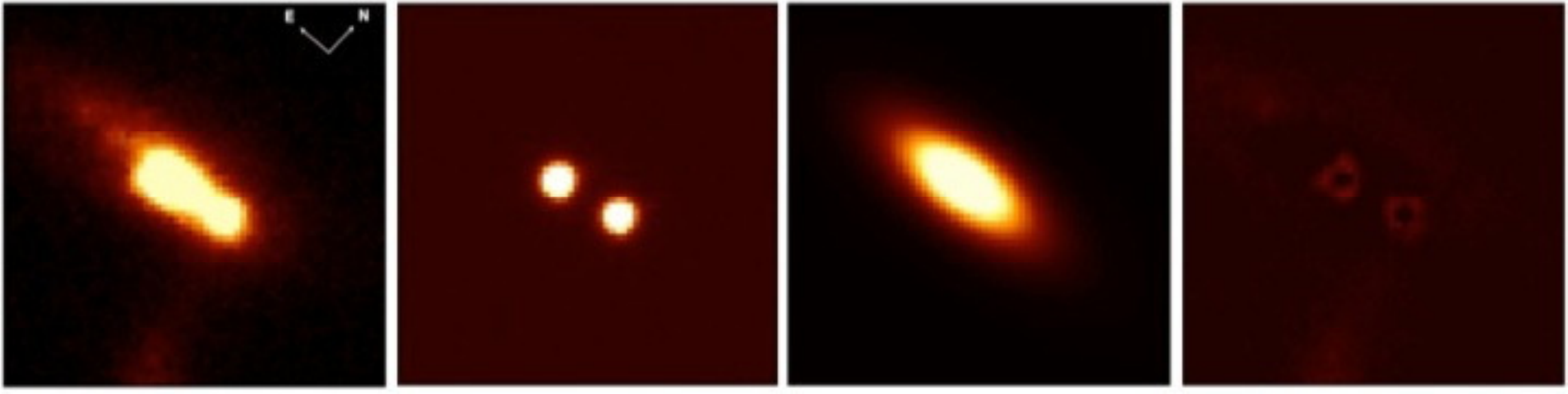}
  \protect\caption[PSF fitting]{PSF fitting of Mrk 19. Figures from left to right show the observed \textit{r}-band image from SDSS DR8, the model image of the two nuclei fitted by \texttt{imfitfits}, the model image of the host galaxy fitted by \texttt{imfitfits}, and the residual image. The image sizes are 25 arcsec $\times$ 25 arcsec. \\(A color version of this figure is available in the online journal.) \label{PSFfitting}}
\end{figure*}

\section{Data analysis}
\label{analysis}
\subsection{PSF fitting}
The analysis of the images is in a first step attempted using the
\texttt{IRAF}\footnote{IRAF is distributed by the National Optical
  Astronomy Observatories operated by the Association of Universities
  for Research in Astronomy, Inc. under cooperative agreement with the
  National Science Foundation.} software and the PSF fitting code
\textit{imfitfits} (\citealt{1998AJ....115.1377M}), which fits a model
convolved with a PSF image. The model is optimized by computing the
$\chi^{2}$ to fit the observed image. In order to determine the
brightness of each individual nucleus, the field images are first
cropped to a size of 64 $\times$ 64 pixels that includes the two nuclei and the host galaxy of each target.
An isolated, bright, but not saturated (less than 15,000 counts at the peak) star
is chosen from the field image and a region of 64 $\times$ 64 pixels
is cut around it. Using this star image as the PSF, we attempt a
two-dimensional-multicomponent fitting of the central region where the two nuclei
lie. 
In principle, we may have attempted a fit based on the canonical galaxy components: one or two galaxy disks, one or two bulges, and two nuclei. However, the spatial resolution is not sufficient to identify the slope changes at the inner central regions and the use of three components yields a degeneracy in the solutions. In addition, it is not clear whether such physical decomposition applies to minor mergers where the two nuclei are present. We thus decide to adopt a less ambitious approach and distinguish only between compact and extended components by using only two components per merger system: two PSFs for the nuclei, and one or two exponential disks for the galaxy or galaxies. This should provide statistically reasonable estimates of the intensities of the compact components and good measurements of the nuclear distances, while reducing the degeneracy as much as possible. 

\begin{figure*}
    \includegraphics[scale=0.33]{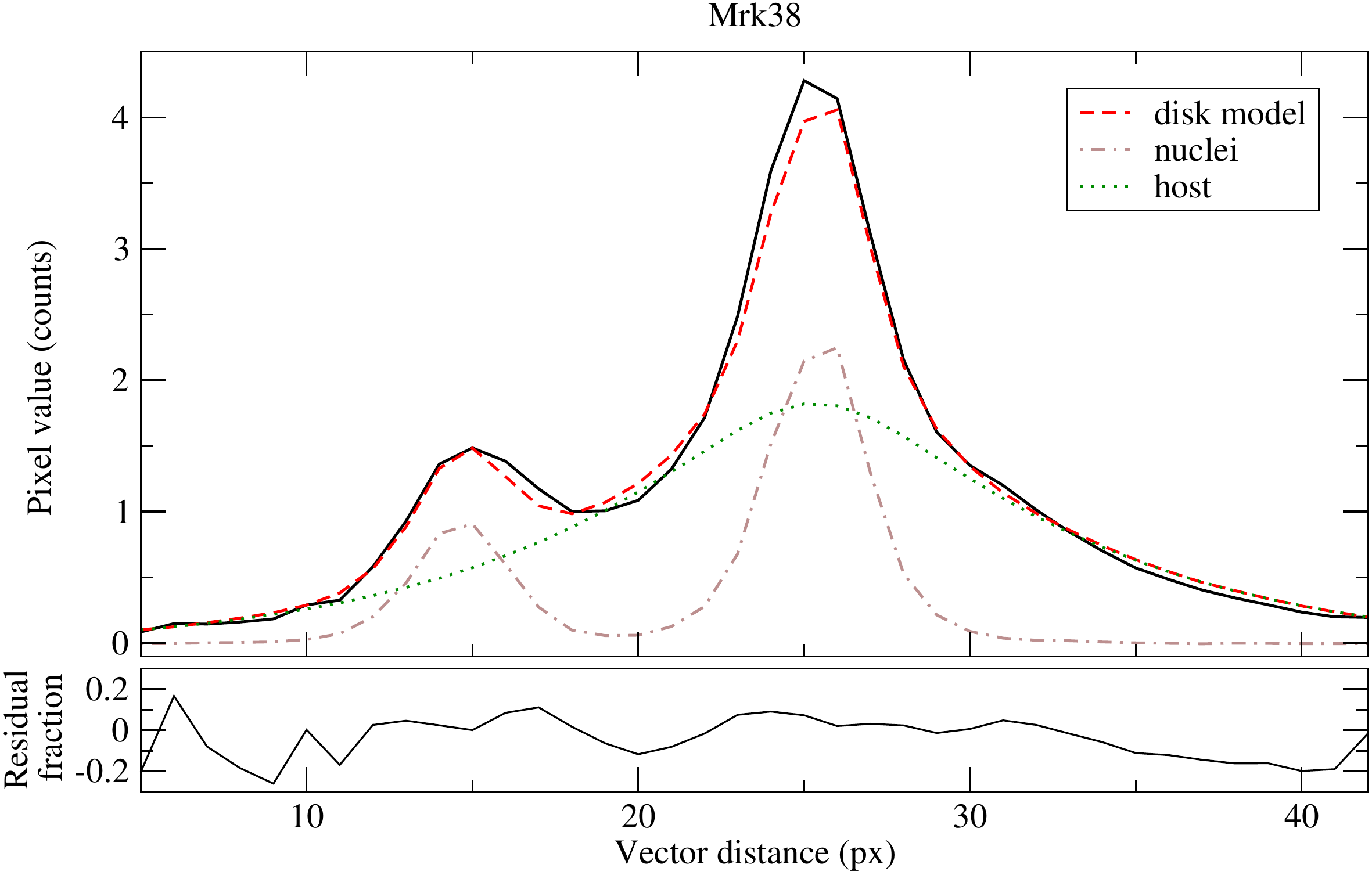}
    \hspace{10pt}
      \includegraphics[scale=0.33]{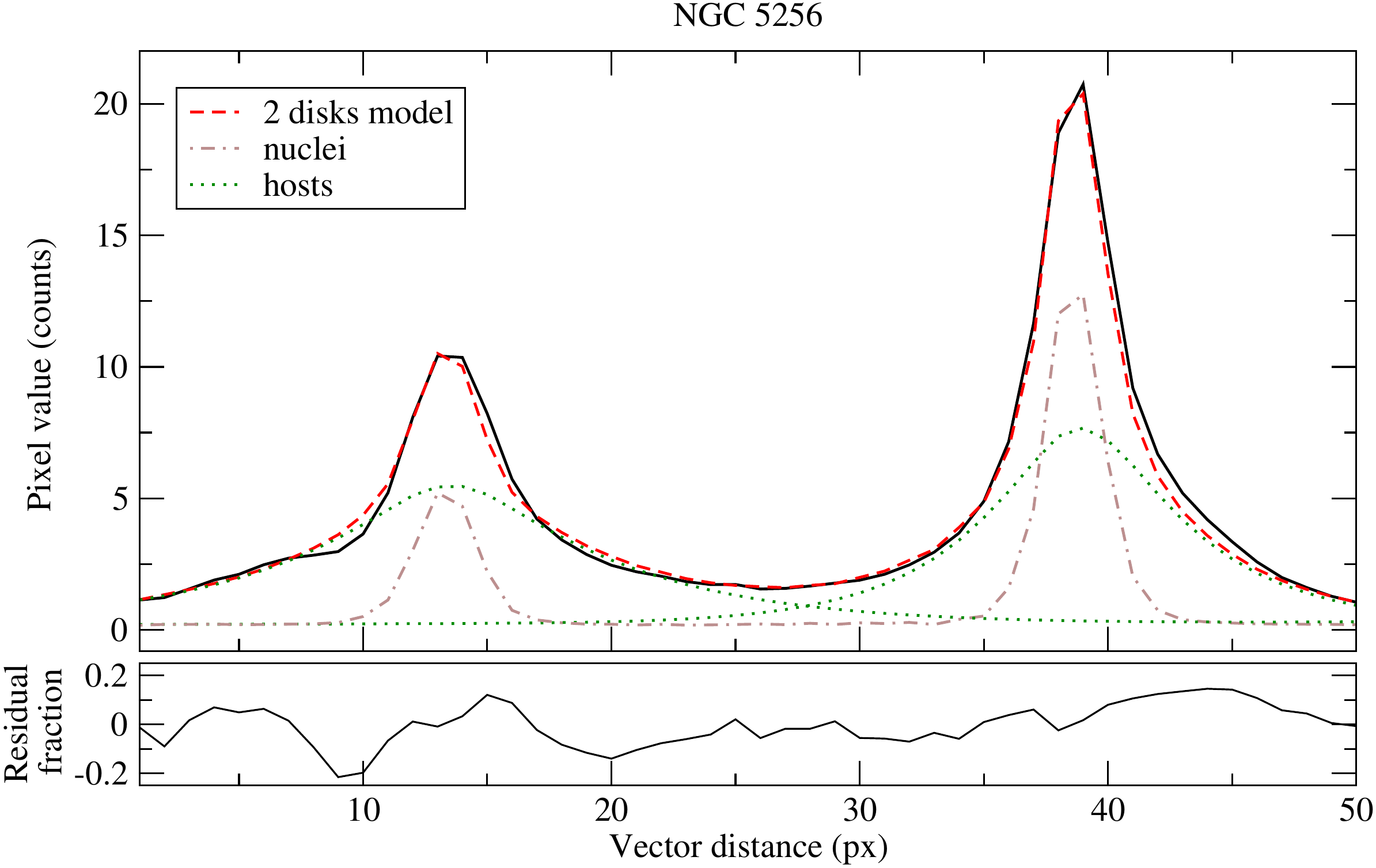}
  \protect\caption[imfitfits fitting]{One-dimensional vector (solid black line) extracted along the position of the nuclei of Mrk 38 (left) and NGC 5256 (right). The two-dimensional \texttt{imfitfits} photometric fit of a disk (for Mrk 38) and 2 disks (for NGC 5256) is shown (red dashed line) decomposed into the host fit (green dotted line) and the fit to the two nuclei (brown dotted-dashed line). The fractional residuals are shown at the bottom of each plot. \\(A color version of this figure is available in the online journal.)\label{fit_imfitfits}}
\end{figure*}

Using this decomposition we are able to perform satisfactory two-dimensional fittings to 50\% of the double nucleus
galaxies in our sample. The uncertainty of the
nuclear intensities fitted using \texttt{imfitfits} can be estimated
from the residual image created after subtraction of the model image
from the data (see Figure~\ref{PSFfitting}). We obtain an uncertainty $\leq$10$\%$, which supports the consistency of the photometric decomposition. 
In one case (Mrk\,22) we need to consider a de
Vaucouleurs profile to fit the host. In order to visually check the
goodness of the fits, we extract a one-dimensional vector along the position of the
nuclei and subtract the model fit from the observed profile to visualize the residuals (see Figure~\ref{fit_imfitfits}).

\begin{figure*}
\includegraphics[scale=0.33]{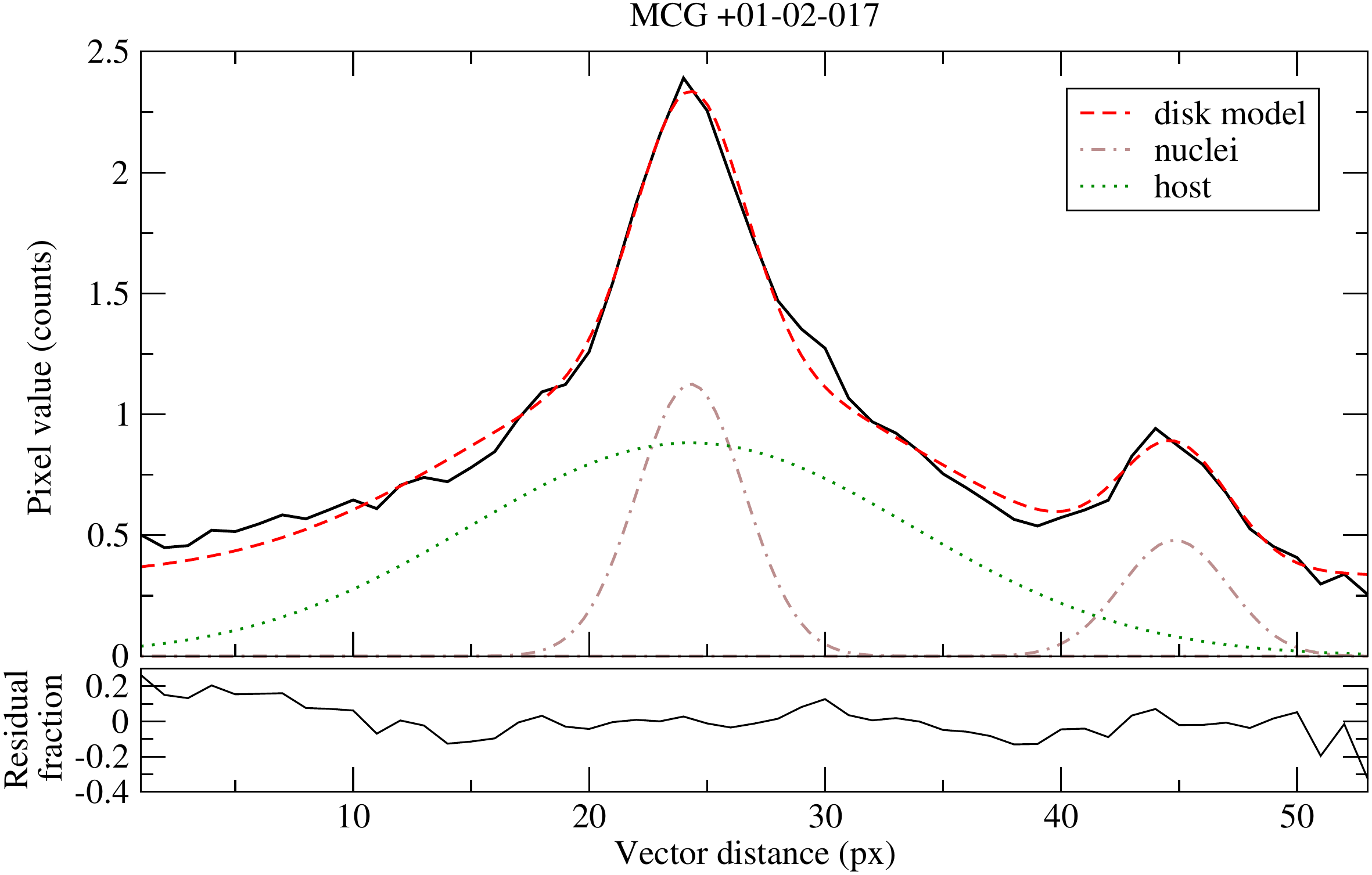}  
    \hspace{10pt}
\includegraphics[scale=0.33]{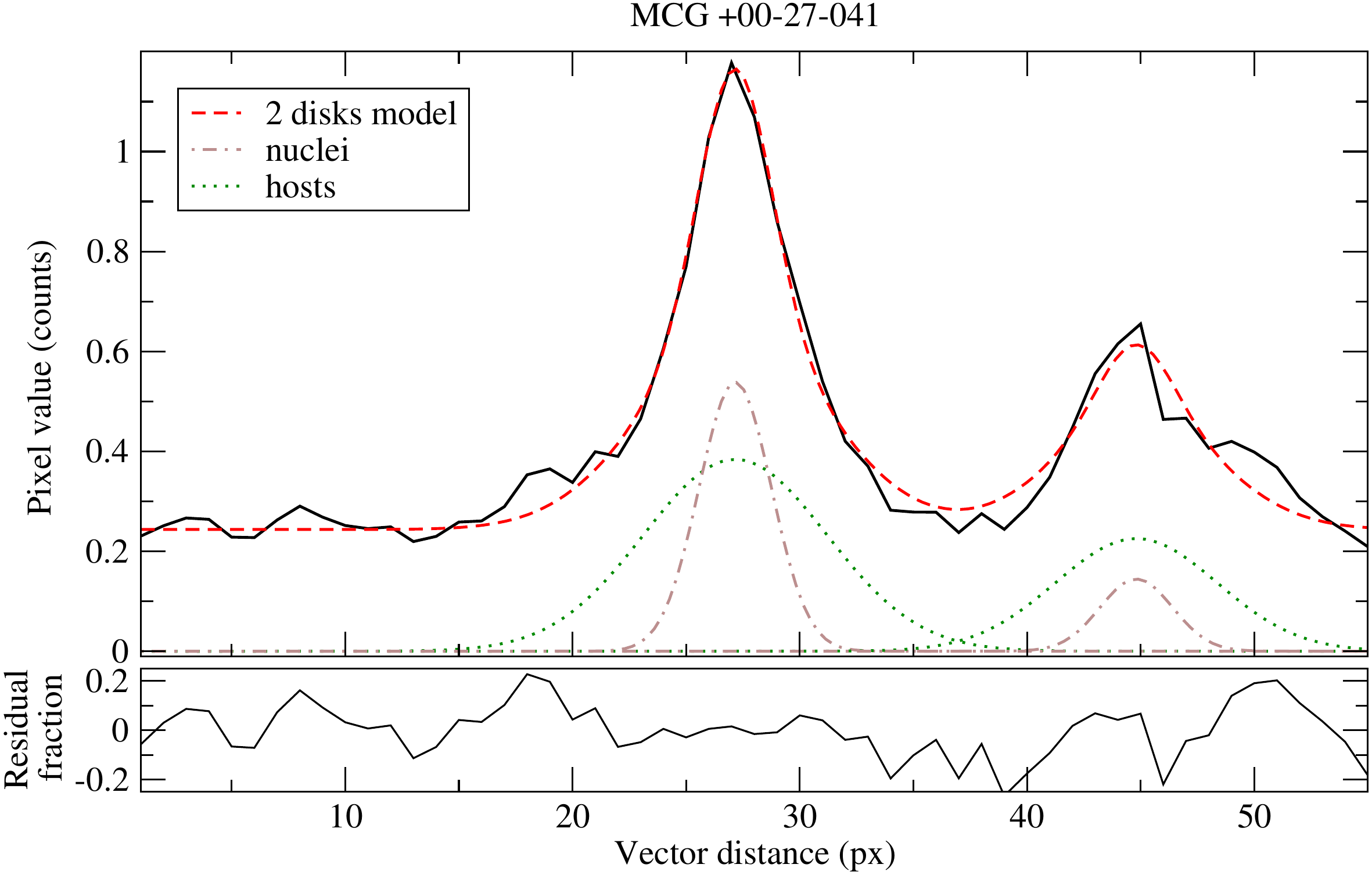}
 \protect\caption[DIPSO fitting]{One-dimensional \texttt{DIPSO} photometric fit of MCG +01-02-017 (left) and MCG +00-27-041 (right). As in the previous plot, the lines plotted are: profile (solid black line), photometric fit (red dashed line), host galaxy fits (green dotted line), and fit of the two nuclei (brown dotted-dashed line). One disk model is needed to fit the host galaxy of MCG +01-02-017, while two disks are required to fit the host of MCG +00-27-041. The fractional residuals are shown at the bottom of each plot. \\(A color version of this figure is available in the online journal.)\label{fit_dipso}}
\end{figure*}

For those galaxies whose luminosity profiles are not well-fitted using a
two-dimensional photometric model, a one-dimensional vector is extracted along the position of
the nuclei using the \texttt{IRAF} task \texttt{pvector}. These one-dimensional
profiles are then fitted with Gaussian components using the Starlink
\texttt{DIPSO} software package (\citealt{howarth}). A Gaussian of
fixed width is fitted to each of the nuclei (the width is constrained
to a value $\simeq$ to the stellar PSF), while a Gaussian of variable
width is used to fit the host galaxy. In some cases, two Gaussians are required to fit two
host galaxy disks (see Figure ~\ref{fit_dipso}). The
errors of the nuclear intensities provided by \texttt{DIPSO} are
$\leq$20$\%$. 
To explore the systematic errors due to the two
different photometric models used, we compare the two-dimensional fitting of
\texttt{imfitfits} with the one-dimensional results from \texttt{DIPSO} for those galaxies that could be fitted by the two models. The
difference between the two models is going to be a conservative estimate
of the systematic errors induced by the photometric modeling. 
We find that the differences in nuclear intensities are
$\leq$30$\%$.

The error in the relative separation between the two nuclei is derived as the quadratic sum of the astrometric error of each nucleus (provided by the global astrometric precision of the SDSS DR8, which is 0.1 arcsec) and the uncertainty of the photometric fit (we use conservative limits of 10$\%$ for the 
sources fitted with \texttt{imfitfits} and 30$\%$ for the ones fitted with \texttt{DIPSO}).

\subsection{Nuclear and Host Galaxy Luminosities}
The intensities obtained from
\texttt{imfitfits} for each of the two nuclei and host galaxies are converted to magnitudes
using the calibration specified in the header of each SAS
image. We correct these magnitudes
for Galactic dust extinction using the reddening corrections at the
position of each object provided by SDSS (following
\citealt{1998ApJ...500..525S}) and convert them to absolute AB magnitudes (\citealt{1983ApJ...266..713O}). The \textit{r}-band luminosities of the nuclei are then derived using
$L/L_{\odot}=10^{(M-M_{\odot})/2.5}$, 
where $M$ is the absolute magnitude of the target, and $M_{\odot}$ and $L_{\odot}$ are, respectively, the absolute \textit{r}-band magnitude and luminosity of the Sun. Their values, used here, are $M_{\odot}$ = 4.42 and $L_{\odot}$ = 3.85$\times$10$^{26}$ W (e.g., \citealt{1995Sci...268.1640P}; \citealt{1998gaas.book.....B}). 

\section{Results}
\label{results}
The results of the PSF fitting (i.e., magnitudes and luminosities of the primary and
secondary nucleus and the host galaxy, and the relative separation between the nuclei) together with the
name and distances of the galaxies analyzed are listed in
Table~\ref{doublenucleus}. 
Of the 60 objects for which SDSS DR8 images are available, three do not have any spectroscopical identification nor redshift
available and thus their nuclear and host galaxy luminosities could
not be derived. These objects are therefore excluded from the analysis
and are not included in Table~\ref{doublenucleus}. The
fitting codes are not able to find a double nucleus in three
other targets, which are thus also removed from the analysis. The host
galaxy of two other sources, MCG +06-21-031 and MCG +10-21-040, could
neither be fitted by a two-dimensional nor by a one-dimensional model as the SDSS images were
very noisy. These two sources are also excluded from the
analysis. Therefore, the total number of objects included in
Table~\ref{doublenucleus} and considered for further analysis is 52.

\subsection{Extinction}
\begin{figure}[h!]
\vspace{17pt}
\centering
 \includegraphics[scale=0.31]{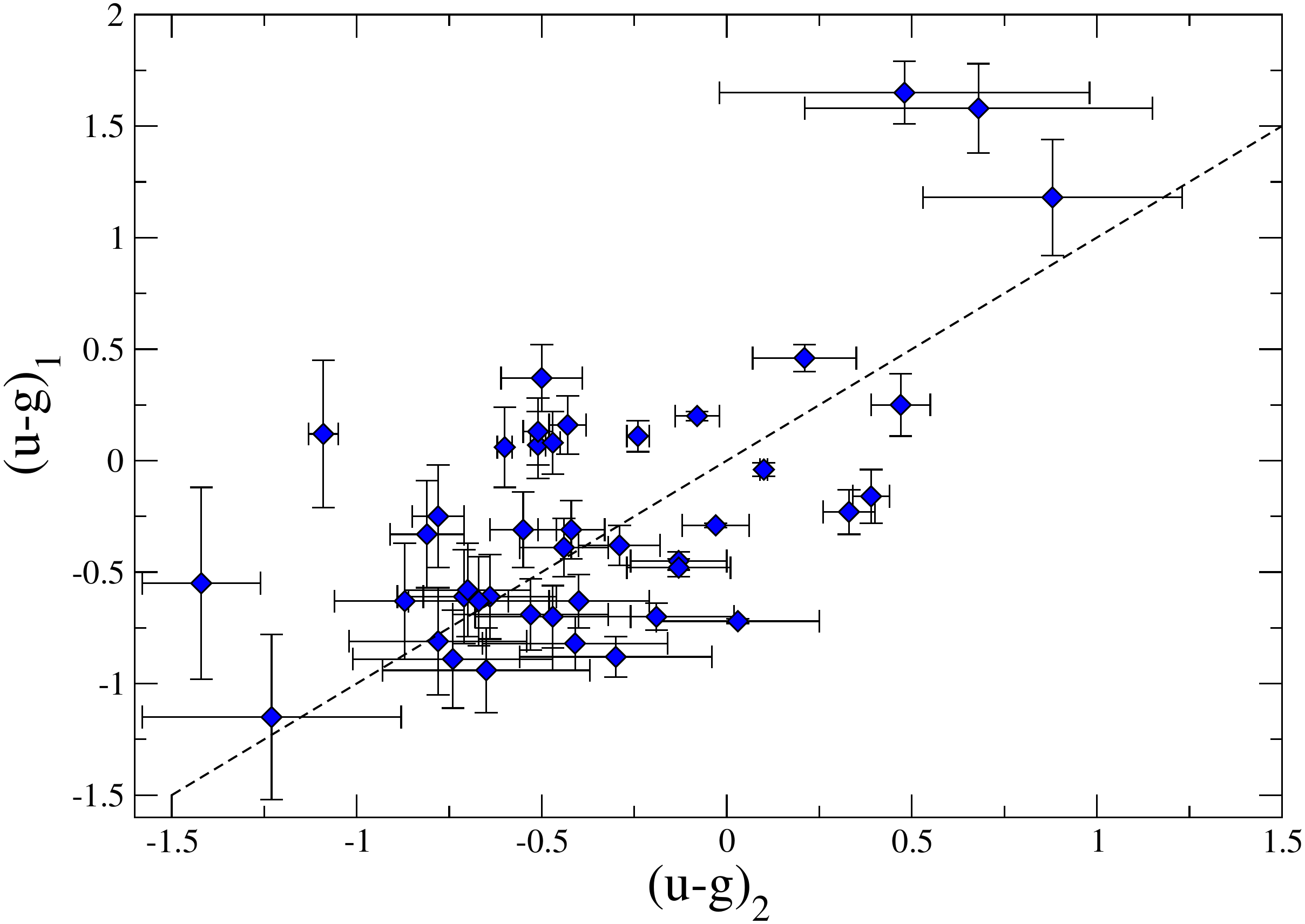}
  \protect\caption[Color-color diagram]{Color--color diagram of  \textit{u -- g} magnitudes of the primary vs. the  \textit{u -- g} magnitudes of the secondary nucleus. A one-to-one correlation indicating no systematic difference in extinction between the two nuclei is also plotted (dashed line).\\(A color version of this figure is available in the online journal.)\label{ug1_ug2}}
\end{figure}

To ensure that the results obtained are not affected by different
degrees of extinction in each of the nuclei, the \textit{u} and
\textit{g}-band images of the 52 target sources in
Table~\ref{doublenucleus} are retrieved from the SDSS DR8 and PSF
fitting with \texttt{imfitfits} is performed in these bands. The
images are first cropped to 64 $\times$ 64 pixels and a point source
model is fitted to each of the nuclei using a stellar image as
PSF. The parameters of the model are the positions and intensities of
each of the nuclei and the background level of the image. The steps are the same as described in Section~\ref{analysis}.  In
the \textit{u} band, \texttt{imfitfits} fails to fit the positions of
the double nuclei in nine of the targets. For the remaining 43 sources,
the fitted fluxes are converted from nanomaggies to magnitudes and the
\textit{u--g} colors of the primary against the secondary nucleus are
plotted in Figure~\ref{ug1_ug2}.  The figure shows that both nuclei seem
to be in general affected by the same extinction, indicating that they both have
the same environment. The results obtained in the previous sections (Table~\ref{doublenucleus})
are thus not affected by differences in the ambient medium in which
the nuclei are embedded.

We also derive the amount of extinction
allowed by the fit of the nuclear luminosities.
Within the 30\% error
of the nuclear intensities, an extinction of $A_{r}\leq0.3$ mag is
allowed. A higher degree of extinction might be present in these
double nucleus galaxies, where nuclear
star-formation might be contributing to the observed nuclear
intensities (see Section~\ref{discussion}).

\subsection{Major and Minor Mergers}
\label{sec:minor}
The target sources were classified as minor merger candidates based on their morphological appearance lacking strong tidal tails (\citealt{2004AJ....128...62G}). 
However, the lack of these features can also indicate that some sources might be in a late stage of a major merger event when the optical morphological distortions have already vanished and are only recognizable in studies of the neutral hydrogen (HI) gas (e.g., \citealt{2006A&A...454..125E}; \citealt{2009MNRAS.400.1749K}).
In the case of very shallow imaging, like in the standard-depth SDSS images analyzed here, the lack of visible morphological distortions results from the very short exposure times (i.e., of only 53.9 s per band), which are insufficient for revealing the tidal features owing to their faintness compared to the host galaxy magnitude. To detect tidal features around mergers (especially for minor mergers) much deeper imaging is required (e.g., deep Stripe82 images; \citealt{2010MNRAS.406..382K}; see also \citealt{2005AJ....130.2647V}; \citealt{2012ApJS..202....8S}). 

To determine what is the stage of the merger event and whether morphological features should be present, we derive the ratio of nuclear luminosities $L_\mathrm{2}/L_\mathrm{1}$, usually adopted to classify sources into major and minor mergers (e.g., \citealt{2012A&A...539A..45L}). We adopt a luminosity ratio $>$1:3 for major mergers. We obtain that only 38\% of the target sources (20 out of 52) qualify as minor mergers, while most of the sources studied (62\%) are classified as major mergers. 
Since these sources exhibit two nuclei, most of these systems must be in the coalescence stage of a major merger in which the nuclei are in the process of merging but have not yet merged and a single galaxy with morphological distortions is formed (e.g., \citealt{2005MNRAS.361..776S}; \citealt{2008MNRAS.391.1137L}). The non-detection of these strong tidal features is thus caused by the low exposure time of the SDSS images. The finding that most of the galaxies studied qualify as major mergers despite their initial morphological classification as minor mergers (\citealt{2004AJ....128...62G}) is of significant importance for studies based on visual classification of galaxies, as many so-called ``minor mergers'' might actually be mis-classifications of major mergers.

\subsection{The Coalescence Stage}
\label{sec:merger}
\begin{figure}[h!]
\centering
\vspace{20pt}
  \includegraphics[scale=0.32]{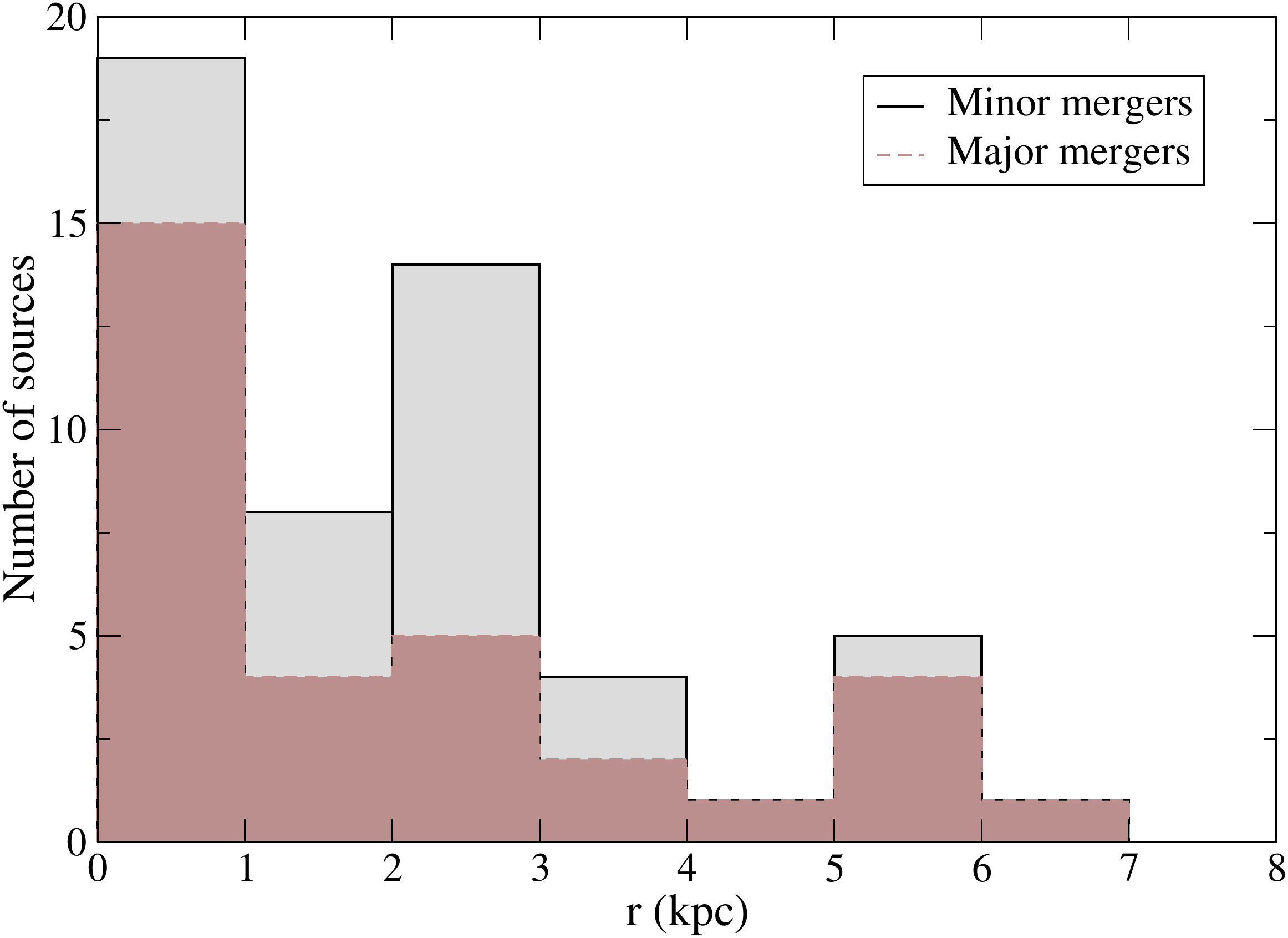}
  \protect\caption[Histogram separation]{Distribution of projected separations between the two nuclei for sources classified as major mergers (dashed line, red bars) and minor mergers (solid line, gray bars). \\(A color version of this figure is available in the online journal.)\label{separation}}
\end{figure}

We find that in 19 out of the 52 sources the two nuclei are separated by a projected distance $\leq 1$ kpc (see Figure~\ref{separation}). A secondary peak in the distribution is observed for objects with nuclear projected separations between 2 and 3 kpc. A total of 91\% of the sources have nuclear projected separations $\leq 4$ kpc, which is in agreement with the results from \cite{2004AJ....128...62G}. As suggested by these authors, the lower number of double nucleus galaxies with larger nuclear separations
may be a selection effect: the greater the distance between the host galaxy and the merging satellite, the less recognizable is the merger system as a double nucleus galaxy. The double nucleus galaxies with the shortest nuclear separations would 
thus belong to an evolved stage of the merger (i.e., the coalescence phase) in which a single galaxy has already been formed.
This is supported by the finding that 65\% (33 out of 51\footnote{The fit of one source required no host galaxy disks.}) of the sources were fitted by one disk host galaxy, which indicates that the disk of the satellite galaxy is no longer recognizable and is in agreement with the finding that 62\% of the target sources qualify as major mergers according to the ratio of nuclear luminosities (see Section~\ref{sec:minor}). 
On the other hand, the main peak of 19 sources with nuclear separation $\leq 1$ kpc (Figure~\ref{separation}) is mainly formed by the major mergers systems, while the secondary peak in the range 2--3 kpc is owing to the peak in the distribution of minor mergers. These results yield further support to the suggestion that those systems with the shortest nuclear distances must be in the coalescence stage of a major merger.

\section{Discussion}
\label{discussion}
\subsection{The Nature of the Nuclei}
Identifying the energy source of the nuclei studied here is not trivial. We have shown that, for the majority of these sources, the scenario of a coalescence phase in a major merger is more plausible rather than that of an ongoing minor merger. For these sources, the presence of a black hole (BH) in each of the two nuclei is expected. This BH can either be actively accreting or be dormant. Conversely, for those sources identified as minor mergers, the secondary ÒnucleusÓ can potentially in addition be either a star-forming region or a stellar cluster. To test these different possibilities, we investigate whether these sources are detected in the mid-IR (MIR, by the \textit{Wide-field Infrared Survey Explorer}, (\textit{WISE}); \citealt{2010AJ....140.1868W}) or the far-UV (by the \textit{Galaxy Evolution Explorer (GALEX)}; \citealt{2005ApJ...619L...1M}). Given the small angular separation of the two nuclei, they cannot be resolved by \textit{WISE} nor \textit{GALEX} in almost all the galaxies of our sample. However, the \textit{WISE} colors of the 52 sources reveal that 2 of them ($\sim$4\%), MCG +00-04-098 and NGC 3758, have AGN-like colors (e.g., \citealt{2012ApJ...753...30S}; \citealt{2012MNRAS.426.3271M}), while the rest occupy the star-formation region of the plot. These results suggest that star formation rather than AGN dominates the emission in the majority of the sample, although the presence of a deeply embedded AGN cannot be excluded. Such highly obscured AGN are expected during the process of disk-galaxy major mergers (e.g., \citealt{1988ApJ...328L..35S}; \citealt{2001ApJ...555..719C}). In that case, higher-resolution MIR observations are required to distinguish between AGN and star formation by separating the nuclear and the extended MIR contribution, as is the case of Mrk 789, which is known to host a Seyfert nuclei (e.g., \citealt{2007AJ....134.2006R}) despite its location outside the AGN locus in the \textit{WISE} color diagram.

The presence of an AGN can be also confirmed by hard X-ray observations (if the X-ray emission is Compton thin) with high angular resolution (with the \textit{Chandra} X-ray satellite) in the case of binary AGN at kpc scales. Several kpc-scale binary AGNs have been found using X-ray observations
(e.g., \citealt{2003ApJ...582L..15K}; \citealt{2005A&A...429L...9G};
\citealt{2006A&A...453..433H}; \citealt{2008MNRAS.386..105B};
\citealt{2011ApJ...737L..19C}; \citealt{2011ApJ...735L..42K}). The
closest separations in an AGN pair have so far been detected in major
mergers of spiral galaxies (NGC\,6240, projected separation $\sim1$
kpc; \citealt{2003ApJ...582L..15K}, and Mrk\,739, projected separation
$\sim3.4$ kpc; \citealt{2011ApJ...735L..42K}), 
setting strong constrains on the co-evolution of galaxies and their
nuclear black holes. The occurrence of minor mergers should be more
common; however, only one AGN pair has been found in a merger of
un-equal mass galaxies (NGC\,3393, projected separation $\sim150$ pc;
\citealt{2011Natur.477..431F}).
In the total sample of 52 double nucleus systems, the PSF fitting
finds 19 sources with projected nuclear separations $\leq 1.0$ kpc, or physical
separations $\leq 1.1$ kpc when considering an average inclination angle
for the host galaxies of $\sim69^{\circ}$ (\citealt{2004AJ....128...62G}). Most of the 19 sub-kpc pairs qualify as major mergers (15 sources), while 4 of the sources are classified as minor merger events (see Section~\ref{sec:merger}). 
We have searched for any \textit{Chandra} X-ray studies or observations available in the literature and the archives of the 52 sources analyzed in this paper. We find that three sources have been studied in detail with \textit{Chandra} and that in two of them (NGC\,5256, \citealt{2007MNRAS.377.1439B}; NGC\,3758, \citealt{2011ApJ...735L..42K}) the presence of a binary AGN is confirmed. The nuclear separations obtained from the SDSS photometric fit (6.2 arcsec for NGC\,3758 and 10.2 arcsec for NGC\,5256, see Table~\ref{doublenucleus}) are consistent with the separations derived from the {\it Chandra} observations. NGC\,5256 qualifies as a major merger according to the ratio of nuclear luminosities, in agreement with the classification of \citealt{2007MNRAS.377.1439B}, while NGC\,3758 is classified as a minor merger. Archival \textit{Chandra} data are available for four more sources, whose analysis is left for future work.

Finally, nuclear accretion within the galaxy pairs could also be probed from their location in emission-line diagnostic diagrams, although, as seen above, star formation can in some cases overwhelm the AGN emission, whose presence can only by definitely confirmed by hard X-ray observations. This is the case, for instance, for NGC\,3758, which was confirmed as a binary AGN from \textit{Chandra} observations although the secondary nucleus does not qualify as an AGN based on its optical emission lines (\citealt{2011ApJ...735L..42K}).

Of particular interest is the case of the presence of a low-mass black hole in the satellite galaxy nucleus of those 20 sources that we identified as minor mergers. Low-mass BHs (or intermediate-mass BHs, IMBHs) are expected in the nuclei of low-mass dwarf galaxies from the BH-mass scaling relationships (e.g., \citealt{2013ApJ...764..184M}). The tidal stripping of merging satellite low-mass galaxies should thus yield the presence of IMBHs in the haloes of galaxies (\citealt{2008ApJ...687L..57V}; \citealt{2010ApJ...721L.148B}). Several IMBH candidates have already been found from X-ray and radio studies of ultraluminous X-ray sources (e.g., \citealt{2009Natur.460...73F}; \citealt{2012Sci...337..554W}; \citealt{2011AN....332..379M}; \citealt{2013MNRAS.436.1546M}a, \citeyear{2013MNRAS.436.3128M}b). However, their presence in the nucleus of captured satellite galaxies still remains an open question.

\subsection{The Nucleus--Host Connection}
\label{sec:AGNhost}
The coalescence phase of a merger is seen in simulations to coincide with the peak of merger-induced star-formation and AGN activity (e.g., \citealt{2005MNRAS.361..776S}; \citealt{2008MNRAS.391.1137L}; \citealt{2008ApJS..175..356H}), therefore an increase in nuclear luminosity is expected in the coalescence stage. 

\begin{figure}[h!]
  \hspace{-20pt}
  \includegraphics[scale=0.42]{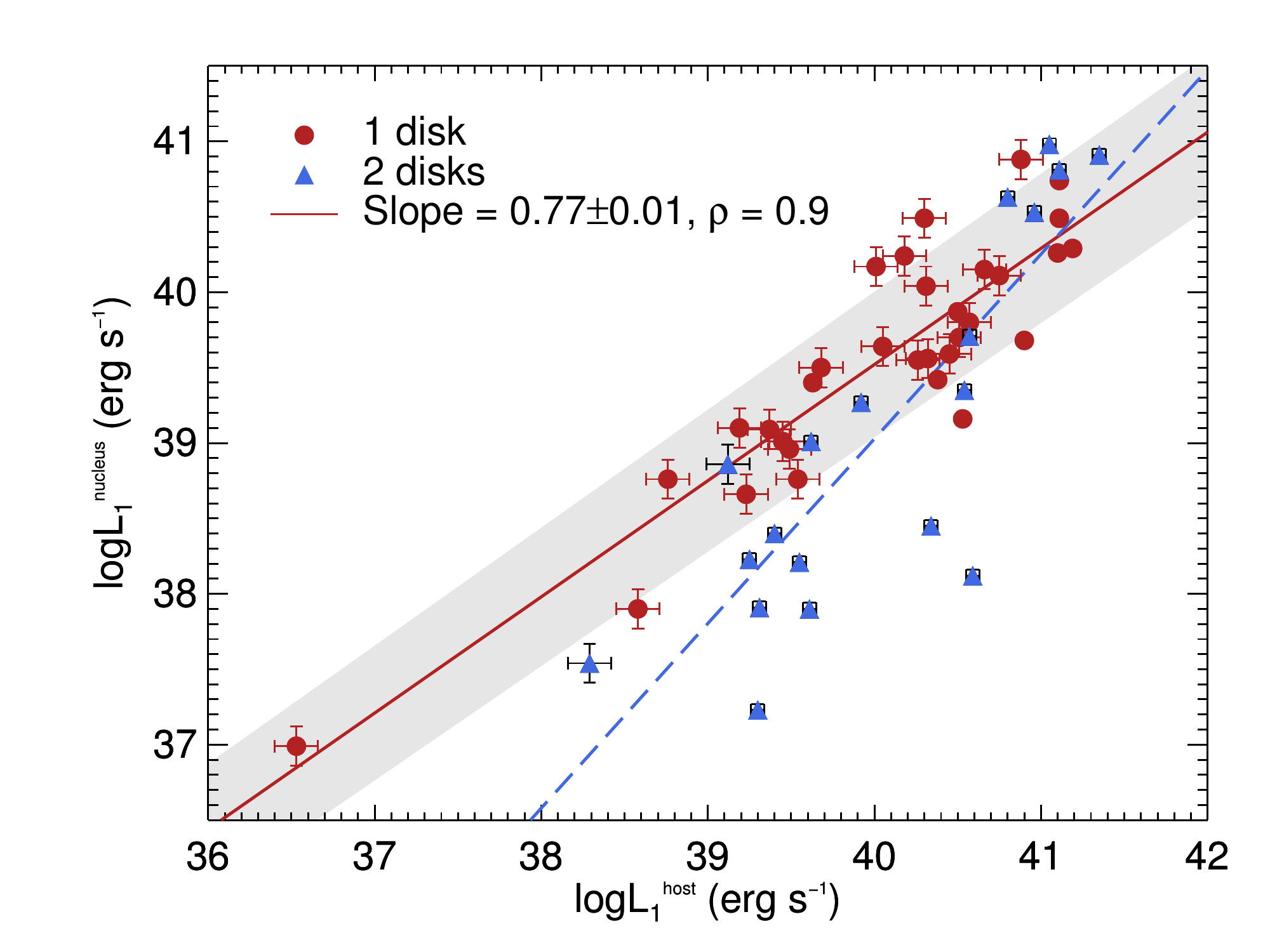}  
  \protect\caption[Nuclear vs. host luminosity]{Luminosity of the primary nucleus vs. host luminosity. Sources with one (red circles) and two (blue triangles) fitted disk host galaxies are plotted. A linear fit to the data for single-disk sources (red solid line) and double-disk sources (blue dashed line) is plotted. We also plot the 1$\sigma$ uncertainty (gray shaded area) of the fitted slope for the single-disk systems. The slope and Pearson's correlation coefficient of this linear fit are also shown. \\(A color version of this figure is available in the online journal.)
 \label{Lnuclear_Lhost}}
\end{figure}

The detection of both galaxies in the double-disk systems implies an earlier stage of the merging process (i.e., between the first passage and the coalescence stages). In these systems some enhancement of the host galaxy luminosity is expected due to galaxy-wide merger-triggered star formation. Conversely, the single-disk systems are at a more evolved stage of the merger (i.e., the coalescence stage), when the two nuclei should have sunk well into the gravitational center of the merging system. At this stage, single-disk systems should exhibit enhanced nuclear luminosities compared to the double-disk galaxies, caused by either the starburst or the AGN merger-triggered activity. We investigate this by plotting the nuclear luminosity of the primary nucleus versus the host galaxy luminosity (Figure~\ref{Lnuclear_Lhost}).

The trends observed for the two sub-samples of sources (i.e., the single-disk and double-disk systems) are significantly different. We perform a linear fit and correlation analysis separately for each sub-sample. We find a slope of $0.77\pm0.01$ for the single-disk systems (red circles in Figure~\ref{Lnuclear_Lhost}) with a Pearson's correlation coefficient, $\rho$, of 0.9. Although our sample is at very low redshifts ($z<0.05$), we still expect some degree of inter-dependence between the host and nuclear luminosities due to redshift. We calculate the partial correlation coefficient considering this redshift dependence, $\rho_{p}$, and find it to be 0.8 for the single-disk sub-sample. Turning now to the double-disk systems (blue triangles in Figure~\ref{Lnuclear_Lhost}), we calculate a slope of $1.22\pm0.18$ with $\rho=0.8$ and $\rho_{p}=0.4$. The large $\sigma$ value of the slope and the low significance of the correlation as indicated by $\rho_{p}$ imply that, unlike for the single-disk sources, the loose correlation observed for the double-disk systems is mainly due to redshift effects. This indicates that the host galaxy luminosity for the double-disk sources does not or only weakly follows the luminosity of the nucleus.

We note that the slope of the linear regression for the single-disk galaxies is steeper than that of the double-disk systems. When considering the systems with two identifiable disks (blue triangles in Figure~\ref{Lnuclear_Lhost}) at luminosities $<10^{41}$ erg s$^{-1}$, we find that they appear to lie exclusively below the trend observed for the single-disk sources. These findings are in line with the theoretical expectations for single-disk systems described above. For the double-disk systems we expect that at later times (and at smaller separations when the two disks will have merged) the nucleus will also brighten up, putting these sources back on the implied trend.

High luminosity ($>10^{41}$ erg s$^{-1}$) sources with two identifiable host galaxies do not show the nuclear luminosity deficit observed for their lower luminosity counterparts, following the trend implied by the single-disk sources (within $\sim 1\sigma$). These sources might be at an earlier phase of the merger, plausibly having yet to experience the enhancement of star-formation and AGN activity. This conclusion is corroborated by the fact that all five sources exhibit rather large separations between their two nuclei ($r>$4 kpc), occupying the high end of the nuclear separation distribution of the whole sample.

As reported in Section~\ref{sec:merger}, 65\% of the sources studied are under the process of a major merger as implied by their nuclear luminosity ratios and number of host galaxies fitted. The result of the major merger of two disk galaxies is a spheroidal galaxy characterized by the presence of a bulge or pseudo-bulge (e.g., \citealt{2006ApJ...650..791C}; \citealt{2010ApJ...715..202H}; \citealt{2012MNRAS.424.1232K}). The evolution of supermassive BHs (SMBHs) at the centers of bulge and pseudo-bulge galaxies has been found to be closely related to host galaxy properties such as the stellar velocity dispersion (e.g.,  \citealt{2000ApJ...539L...9F}; \citealt{2000ApJ...539L..13G}; \citealt{2011MNRAS.412.2211G}) or the bulge luminosity (e.g., $M_\mathrm{BH} \propto L_\mathrm{bulge}^{0.90\pm0.11}$,\citealt{2002ApJ...565..762W}; $M_\mathrm{BH} \propto L_\mathrm{bulge}^{0.80\pm0.09}$, \citealt{2009ApJ...694L.166B}). 

As shown in Figure~\ref{Lnuclear_Lhost}, the linear regression analysis shows a clear positive correlation between $L_\mathrm{1}$ and $L_\mathrm{host1}$ of slope 0.77$\pm$0.01 for the single-disk systems but not for the double-disk galaxies. Given the presence of an actively accreting SMBH within each nucleus, the correlation between the luminosity of the primary nucleus and the host galaxy luminosity, although shallower, seems to be analogous to the $M_\mathrm{BH} \propto L_\mathrm{bulge}^{0.9}$ correlation of scatter 0.1 dex derived both for inactive and active galaxies (\citealt{2002ApJ...565..762W}; see also \citealt{2003ApJ...589L..21M}; \citealt{2009ApJ...694L.166B}). This correlation is not expected in disk galaxies but only in systems with bulges or pseudo-bulges. Its emergence for merging systems with one host disk, but not with two host disks, can be explained by them qualifying as major mergers in which the putative SMBHs are or will go on to form a bulge or pseudobulge and eventually fall on the $M_\mathrm{BH}\propto L_\mathrm{bulge}$ correlation. 


\section{Conclusions}
\label{conclusions}
In this paper, we have performed a photometric PSF fitting for a sample of double nucleus disk galaxies candidates for minor merger. This has enabled estimating the luminosity of the primary nucleus, the secondary nucleus, and the host galaxy of 52 double nucleus disk galaxies and the distance between the two nuclei.
Despite the initial classification of these sources as minor mergers, we find that most of the targets qualify as major mergers based on the ratio of nuclear luminosities (32 out of 52 sources) and on the finding that 65\% of the host galaxies are fitted by one exponential disk. These sources must therefore be in a more evolved stage of the merger (i.e., the coalescence stage) than the double-disk systems. We find that $\sim$90\% of the sources studied have projected nuclear separations $\leq 4$ kpc, of which in 19 sources the two nuclei are separated by $\leq 1$ kpc. There are very few confirmed cases of kpc-scale binary AGN, therefore these 19 sources are potential binary AGN candidates to be targeted by high-resolution X-ray observations. Finally, we present evidence for the systems fitted with one host to show enhancement of their nuclear luminosity as expected from simulations. This is possibly caused by the peak of the merger-triggered star-formation and AGN activity.

\section*{Acknowledgments}
The authors are grateful for the suggestions of the anonymous referee that helped to significantly improve the manuscript.
The authors are very thankful to Joe Mazzarella and Richard Davies for insightful discussions.
M.M. was supported for this research through a stipend from the
International Max Planck Research School (IMPRS) for Astronomy and Astrophysics at the Universities of Bonn and Cologne. M.K. was supported for this research by the National Research Foundation of Korea (NRF) grant, No. 2008-0060544, funded by the Korea government (MSIP). Part of this work was supported by the COST Action MP0905 ``Black Holes in a Violent Universe".

\bibliographystyle{mn2e} 
\bibliography{referencesALL}

\begin{deluxetable}{lcccccccccc}
\tablewidth{0pt}
\tablecaption{Double nucleus galaxies\label{doublenucleus}}
\tablehead{Name & $D_\mathrm{L}$ &   $m_\mathrm{r, host}$     & log\,$L_\mathrm{host}$ & $m_\mathrm{r, 1}$ & $m_\mathrm{r, 2}$  & log\,$L_\mathrm{1}$ & log\,$L_\mathrm{2}$ &    $r$ 	&       $r$ 	   & Hosts\\
	& (Mpc)    &  	&	(erg s$^{-1}$)        	&                      &                            & 		(erg s$^{-1}$)	      & (erg s$^{-1}$)	 	&    ($^{\prime\prime}$)           &    (kpc)  &      \\
(1) & (2) & (3) & (4) & (5) & (6) & (7) &(8) & (9) & (10) & (11)	\\ }
\startdata
MCG +00-04-098	&	228	&	22.98	&	40.88	&	22.98	&	24.08	&	40.88	&	40.44	&	2.1	&	2.1	&	1	\\
MCG +00-12-073	&	68	&	21.14	&	40.57	&	23.28	&	23.19	&	39.71	&	39.75	&	17.1	&	5.5	&	2	\\
MCG +00-27-041	&	83	&	25.19	&	39.12	&	25.84	&	27.26	&	38.86	&	38.29	&	7.1	&	2.7	&	2	\\
MCG +01-02-017	&	173	&	22.92	&	40.66	&	24.19	&	25.12	&	40.15	&	39.78	&	8.2	&	6.4	&	1	\\
MCG +01-02-045	&	67	&	24.09	&	39.37	&	24.78	&	26.24	&	39.09	&	38.51	&	4.9	&	1.5	&	1	\\
MCG +01-32-049	&	18	&	21.34	&	39.30	&	26.52	&	25.71	&	37.23	&	37.56	&	7.6	&	0.6	&	2	\\
MCG +02-31-088	&	39	&	23.24	&	39.23	&	24.69	&	26.02	&	38.66	&	38.12	&	12.4	&	2.3	&	1	\\
MCG +02-32-078	&	12	&	27.50	&	36.53	&	26.36	&	26.81	&	36.99	&	36.81	&	8.2	&	0.5	&	1	\\
MCG +05-06-015	&	79	&	21.54	&	40.53	&	24.97	&	23.77	&	39.16	&	39.64	&	5.1	&	1.9	&	1	\\
MCG +06-07-20	&	119	&	23.75	&	40.01	&	23.33	&	25.77	&	40.17	&	39.20	&	4.5	&	2.4	&	1	\\
MCG +07-29-061	&	111	&	21.36	&	40.90	&	24.41	&	26.30	&	39.68	&	38.92	&	4.5	&	2.3	&	1	\\
MCG +10-19-089	&	30	&	24.32	&	38.58	&	26.02	&	27.00	&	37.90	&	37.51	&	6.6	&	0.9	&	1	\\
Mrk 19	&	59	&	21.30	&	40.38	&	23.69	&	23.86	&	39.42	&	39.35	&	4.7	&	1.3	&	1	\\
Mrk 22	&	24	&	21.36	&	39.55	&	24.71	&	28.31	&	38.21	&	36.77	&	4.5	&	0.5	&	2\tablenotemark{a}	\\
Mrk 35	&	15	&	21.34	&	39.19	&	21.56	&	22.74	&	39.10	&	38.63	&	4.0	&	0.3	&	1	\\
Mrk 38	&	155	&	21.35	&	41.19	&	23.62	&	24.94	&	40.29	&	39.76	&	4.4	&	3.1	&	1	\\
Mrk 66	&	89	&	22.34	&	40.32	&	24.25	&	26.06	&	39.56	&	38.83	&	5.9	&	2.4	&	1	\\
Mrk 104	&	33	&	21.76	&	39.68	&	22.21	&	22.75	&	39.50	&	39.29	&	6.2	&	1.0	&	1	\\
Mrk 116	&	12	&	-	&	-	&	22.45	&	23.14	&	38.56	&	38.28	&	5.4	&	0.3	&	0\tablenotemark{b}	\\
Mrk 147	&	100	&	22.64	&	40.30	&	22.15	&	22.04	&	40.49	&	40.54	&	5.0	&	2.3	&	1	\\
Mrk 153	&	36	&	21.35	&	39.92	&	22.98	&	23.84	&	39.27	&	38.92	&	4.5	&	0.8	&	2	\\
Mrk 212	&	99	&	21.35	&	40.80	&	21.77	&	22.86	&	40.63	&	40.20	&	11.8	&	5.4	&	2	\\
Mrk 219	&	47	&	21.08	&	40.26	&	22.85	&	23.56	&	39.55	&	39.27	&	11.4	&	2.5	&	1	\\
Mrk 224	&	15	&	23.56	&	38.29	&	25.44	&	26.08	&	37.54	&	37.28	&	3.9	&	0.3	&	2	\\
Mrk 306	&	73	&	21.28	&	40.57	&	23.21	&	25.00	&	39.80	&	39.08	&	4.8	&	1.6	&	1	\\
Mrk 365	&	68	&	19.76	&	41.11	&	21.32	&	24.13	&	40.49	&	39.36	&	7.2	&	2.3	&	1	\\
Mrk 423	&	140	&	21.34	&	41.11	&	22.09	&	24.86	&	40.81	&	39.70	&	9.2	&	5.9	&	2	\\
Mrk 544	&	94	&	22.47	&	40.31	&	23.15	&	23.51	&	40.04	&	39.90	&	5.5	&	2.4	&	1	\\
Mrk 553	&	58	&	21.34	&	40.34	&	26.06	&	25.65	&	38.45	&	38.62	&	7.3	&	2.0	&	2	\\
Mrk 712	&	68	&	21.30	&	40.50	&	22.89	&	24.29	&	39.87	&	39.31	&	4.9	&	1.6	&	1	\\
Mrk 721	&	140	&	22.24	&	40.75	&	23.83	&	23.74	&	40.11	&	40.15	&	5.4	&	3.4	&	1	\\
Mrk 729	&	186	&	21.35	&	41.35	&	22.46	&	23.00	&	40.91	&	40.69	&	6.2	&	5.1	&	1	\\
Mrk 731	&	25	&	21.32	&	39.62	&	22.84	&	22.51	&	39.01	&	39.14	&	4.0	&	0.5	&	2	\\
Mrk 777	&	134	&	22.74	&	40.51	&	24.76	&	26.70	&	39.70	&	38.93	&	4.6	&	2.8	&	1	\\
Mrk 789	&	137	&	21.30	&	41.11	&	22.21	&	23.35	&	40.74	&	40.29	&	4.7	&	2.9	&	1	\\
Mrk 799	&	42	&	21.39	&	40.05	&	22.42	&	23.26	&	39.64	&	39.30	&	20.8	&	4.2	&	1	\\
Mrk 930	&	71	&	22.21	&	40.18	&	22.06	&	22.73	&	40.24	&	39.97	&	4.8	&	1.6	&	1	\\
Mrk 1114	&	137	&	21.32	&	41.10	&	23.43	&	24.70	&	40.26	&	39.75	&	4.2	&	2.6	&	1	\\
Mrk 1134	&	66	&	21.11	&	40.54	&	24.09	&	26.40	&	39.35	&	38.43	&	2.7	&	0.8	&	2	\\
Mrk 1230	&	26	&	23.60	&	38.76	&	23.59	&	24.42	&	38.76	&	38.43	&	2.2	&	0.3	&	1	\\
Mrk 1263	&	25	&	21.31	&	39.61	&	25.57	&	25.04	&	37.90	&	38.12	&	7.2	&	0.8	&	2	\\
Mrk 1307	&	20	&	21.34	&	39.40	&	23.85	&	24.69	&	38.40	&	38.06	&	8.3	&	0.8	&	2	\\
Mrk 1341	&	21	&	21.25	&	39.49	&	22.58	&	23.80	&	38.96	&	38.47	&	14.4	&	1.4	&	1	\\
Mrk 1431	&	93	&	22.12	&	40.45	&	24.25	&	25.72	&	39.59	&	39.01	&	4.8	&	2.1	&	1	\\
NGC 3049	&	25	&	21.28	&	39.63	&	21.84	&	24.14	&	39.40	&	38.48	&	2.8	&	0.3	&	1	\\
NGC 3758	&	130	&	21.32	&	41.05	&	21.51	&	23.09	&	40.98	&	40.35	&	6.2	&	3.7	&	2	\\
NGC 3773	&	18	&	21.08	&	39.45	&	22.18	&	22.32	&	39.01	&	38.95	&	3.2	&	0.3	&	1	\\
NGC 4509	&	17	&	21.36	&	39.25	&	23.92	&	26.27	&	38.23	&	37.29	&	8.3	&	0.7	&	2	\\
NGC 5058	&	17	&	21.30	&	39.31	&	24.80	&	25.37	&	37.91	&	37.68	&	2.5	&	0.2	&	2	\\
NGC 5256	&	119	&	21.35	&	40.96	&	22.44	&	23.57	&	40.53	&	40.08	&	10.2	&	5.6	&	2	\\
NGC 5860	&	77	&	21.34	&	40.59	&	27.50	&	24.02	&	38.12	&	39.51	&	9.1	&	3.3	&	2	\\
NGC 7468	&	24	&	21.39	&	39.54	&	23.35	&	24.45	&	38.76	&	38.32	&	7.7	&	0.9	&	1	\\
\enddata
\tablecomments{(1) Object name; (2) luminosity distance (from NED); (3 and 4) \textit{r}-band apparent magnitude and luminosity, respectively, of the primary host galaxy from the photometric fit; (5 and 6) \textit{r}-band apparent magnitude of primary and secondary nucleus, respectively, from the photometric fit; (7 and 8) \textit{r}-band primary and secondary nuclear luminosity, respectively; (9 and 10) projected separation in arcsec and kpc, respectively; (11) number of disk host galaxies fitted. $^\mathrm{a}$ A de Vaucouleurs profile is needed to fit the host. $^\mathrm{b}$No galaxy disk is fitted, which can be explained by the morphology of the host galaxy as a blue compact irregular (I Zw 18).}
\end{deluxetable}

\clearpage

\end{document}